\documentclass[twocolumn,preprintnumbers,amsmath,amssymb,nobibnotes,nofootinbib,superscriptaddress]{revtex4}
\usepackage{graphicx}
\usepackage{dcolumn}
\usepackage[usenames,dvipsnames]{color}
\RequirePackage[colorlinks=true,urlcolor=blue,anchorcolor=blue,citecolor=blue,filecolor=blue,
               linkcolor=blue,menucolor=blue,linktocpage=true,pdfproducer=medialab]{hyperref}
\usepackage{amsmath}
\usepackage{mathrsfs}
\usepackage{slashed}
\usepackage{footnote}

\def\cA{{\cal A}}

\def\cH{{\cal H}}
\def\cO{{\cal O}}

\def\Tr{{\rm Tr}}

\def\be{\begin{equation}}
\def\ee{\end{equation}}
\def\beq{\begin{equation}}
\def\eeq{\end{equation}}
\def\bc{\begin{center}}
\def\ec{\end{center}}
\def\bea{\begin{eqnarray}}
\def\eea{\end{eqnarray}}

\newcommand{\eV}{\;\text{eV}}

\newcommand{\mean}[1]{\langle#1\rangle}

\newcommand{\ep}{\epsilon}
\newcommand{\diag}{\mathrm{diag}}

\newcommand{\mbD}{\mathbf{D}}
\newcommand{\mbTh}{\mathbf{\Theta}}

\newcommand{\lhood}{\mathcal{L}}
\newcommand{\ev}{\mathcal{Z}}

\newcommand{\mcB}{\mathcal{B}}

\newcommand{\df}{{\rm d}}
\newcommand{\qu}[1]{``#1''}
\newcommand{\eref}[1]{Eq.~(\ref{#1})}

\def\T2K{{\sc T2K }}

\def\MN{{\sc MultiNest}}

\def\L{{\rm L}}
\def\R{{\rm R}}

\newcommand{\refcite}[1]{Ref.~\cite{#1}}
\newcommand{\figref}[1]{Fig.~\ref{#1}}
\newcommand{\tabref}[1]{Tab.~\ref{#1}}
\newcommand{\secref}[1]{Sec.~\ref{#1}}

\newcommand{\blue}[1]{\color{blue} #1 \color{black}}

\begin{document}
%
%

\preprint{\blue{RM3-TH/14-2}}
\preprint{\blue{FTUAM-14-5}}
\preprint{\blue{IFT-UAM/CSIC-14-015}}

\title{\boldmath
A Bayesian comparison of $U(1)$ lepton flavour models
\unboldmath}

\author{Johannes Bergstr\"om}
\email{bergstrom@ecm.ub.edu}
\affiliation{
Departament d'Estructura i Constituents de la Mat\`eria and Institut de Ciencies
del Cosmos,
Universitat de Barcelona, Diagonal 647, E-08028 Barcelona, Spain}

\author{Davide Meloni}
\email{meloni@fis.uniroma3.it}
\affiliation{Dipartimento di Matematica e Fisica, Universit\`a degli Studi Roma Tre,\\ 
Via della Vasca Navale 84, 00146 Roma, Italy}

\author{Luca Merlo}
\email{luca.merlo@uam.es}
\affiliation{Istituto de F\'isica Te\'orica UAM/CSIC and Departamento de F\'isica Te\'orica,\\
Universidad Aut\'onoma de Madrid, Cantoblanco, 28049 Madrid, Spain}

\begin{abstract}
Whether the anarchical ansatz or more symmetric structures best describe the neutrino parameters is a long standing question that underwent a revival of interest after the discovery of a non-vanishing reactor angle and the indication of a non-maximal atmospheric angle. In this letter, a Bayesian statistical approach is adopted in order to analyse and compare the two hypotheses within the context of $U(1)$ flavour models. We study the constraints on individual model parameters and perform model comparison: the results elect constructions with built-in hierarchies among the matrix elements as preferred over the anarchical ones, with values of the evidence that depends slightly on whether the $U(1)$ charges are also considered as free parameters or not, and on the priors used.

\end{abstract}
\maketitle

%
%
\section{Introduction}

The observation of a non-vanishing reactor angle $\theta_{13}$ represents an important achievement towards a more detailed understanding of the lepton flavour sector of the Standard Model. Results on $\theta_{13}$ from reactor and accelerator experiments T2K~\cite{Abe:2011sj,Abe:2013fuq,Abe:2013hdq}, MINOS~\cite{Adamson:2011qu,Adamson:2013ue,Adamson:2013whj}, DOUBLE CHOOZ~\cite{Abe:2011fz,Abe:2012tg}, Daya Bay~\cite{An:2012eh,An:2013zwz} and RENO~\cite{Ahn:2012nd} have been considered in global fits to the Pontecorvo-Maki-Nakawaga-Sakata (PMNS) matrix~\cite{Fogli:2011qn,Schwetz:2011zk,Machado:2011ar,Tortola:2012te,Fogli:2012ua,GonzalezGarcia:2012sz,Capozzi:2013csa}. Interestingly, there is at the moment a controversial indication for deviations from the maximal value of the atmospheric angle $\theta_{23}$, that is advocated in Refs.~\cite{Fogli:2012ua,Capozzi:2013csa}, but not confirmed in Refs.~\cite{GonzalezGarcia:2012sz}.

The discovery of the non-vanishing reactor angle and the indication of a non-maximal atmospheric angle have a deep impact on flavour model building. Indeed, models based on discrete symmetries, that dominated the flavour scenario in the past years for their ability to describe in first approximation specific mixing patters with $\theta_{13}= 0^\circ$ and $\theta_{23}= 45^\circ$~\cite{Altarelli:2004za,Mohapatra:2006gs,Grimus:2006nb,GonzalezGarcia:2007ib,Altarelli:2009wt,Altarelli:2010gt}, need now some adjustment. A few strategies have been followed: introduction of additional parameters in preexisting minimal models; implementation of features that allow next order corrections only in specific directions in the flavour space; search for alternative mixing patterns or flavour symmetries that lead already in first approximation to $\theta_{13}\neq 0^\circ$ and $\theta_{23}\neq 45^\circ$ (see for example the reviews in Refs.~\cite{Grimus:2011fk,Altarelli:2012ss,Bazzocchi:2012st,Morisi:2012fg,King:2013eh,Altarelli:2013eya,King:2014nza} and references therein). In other words, the latest neutrino data can indeed be described in the context of discrete symmetries, but at the prize of fine-tunings and/or eccentric mechanisms. 

This suggests to investigate approaches alternative to discrete flavour symmetries: from models based on continuous symmetries such as $SU(3)$ (i.e. Refs.~\cite{King:2001uz,King:2003rf,deMedeirosVarzielas:2006fc,Davidson:2006bd,Grinstein:2006cg,Alonso:2011jd,Alonso:2012fy,Alonso:2013mca,Alonso:2013nca}) or the simplest $U(1)$ (e.g., Refs.~\cite{Altarelli:2000fu,Altarelli:2002sg,Buchmuller:2011tm,Altarelli:2012ia,Ding:2012wh}), to models where no symmetry at all is acting on the neutrino sector 
(e.g., Refs.~\cite{Hall:1999sn,Haba:2000be,deGouvea:2003xe,deGouvea:2012ac}). 
The latter go under the name of \emph{anarchical} models, which have been characterised as models in which the coupling constants and mass matrices are \qu{random} numbers drawn from an invariant probability distribution.  
It has been claimed that such matrices generically prefer large mixings~\cite{Hall:1999sn,Haba:2000be,deGouvea:2003xe} and that the observed sizable deviation from a zero reactor angle seems to favour anarchical models when compared to other more symmetric constructions \cite{deGouvea:2012ac}. However, as discussed in \refcite{Espinosa:2003qz} for the case of neutrino mass matrices, how much a large value of a parameter is preferred can depend strongly on the definition of \qu{preferred} and of \qu{large}.

It has been suggested in Ref.~\cite{Altarelli:2012ia} that the performances of anarchical models in reproducing the 2012 neutrino data are worse than those of models constructed upon the $U(1)$ flavour symmetry.  The analysis in Ref.~\cite{Altarelli:2012ia} is based on the fact that anarchy can be formulated in a $U(1)$ context, giving no charges to the left-handed fields, but non-zero $U(1)$ charges to the right-handed ones in order to describe the charged lepton mass hierarchy. This indeed allows a consistent comparison between anarchical and hierarchical models. The constructions with built-in hierarchies among the matrix elements considered in Ref.~\cite{Altarelli:2012ia}, that resulted to be favoured with respect to anarchical models, have been chosen due to particular phenomenological features that lead to a good description of the data. On the other hand, these models have not been shown to be necessarily the best ones available and other $U(1)$ models could provide an even better description of the data. The method of analysis used in Ref.~\cite{Altarelli:2012ia} treated all mass matrix elements as random complex numbers with modulus of order one and the models were judged according to the \emph{success probability}, i.e., the fraction of the generated points which satisfied some 
experimentally motivated cuts. This kind of analysis allows to naively estimate the relative success of one model with respect to the others, but it is not based on, nor motivated by, any established method of statistical analysis, and hence the results cannot be given any proper statistical interpretation. However, some aspects of that method are rather good approximations of some procedures having well-defined meanings within Bayesian inference. In comparison, a standard $\chi^2$-analysis of these models is not possible since they can all fit the data equally well for any values of the interesting parameters. The main objective of this work is to make a statistically principled analysis of the models considered in Ref.~\cite{Altarelli:2012ia}, to make a systematic search for even better models, and to make the appropriate generalisations, allowing for a meaningful comparison between anarchy and hierarchy in the form of $U(1)$ models. Furthermore, data on the masses of the charged leptons will be included in the analysis. This is important since this data gives constraints on the model parameters -- constraints that must be consistent with neutrino oscillation data -- which will results in a large impact on the final results.

In the following, basics concepts of Bayesian inference are briefly summarized in \secref{Sect:Bayesian}. Section \ref{u1mod} is devoted to the presentation and analysis of specific flavour models, where the lepton $U(1)$ charges are chosen a priori due to particular features of the Yukawa matrices. In \secref{sec:generalised_models} the lepton charges are instead treated as free parameters in the Bayesian analysis. 
We discuss future prospects to further distinguish among the considered models in \secref{Sect:Future} and present the conclusions in \secref{Conclusions}.

%
%
\section{Bayesian Inference}
\label{Sect:Bayesian}
In the Bayesian interpretation, probability is associated with the \emph{plausibility} or \emph{credibility} of a proposition.
Perhaps the main goal of science is to infer which model or hypothesis best describes a certain set of collected data. Also, these models should preferably be \qu{simple} or \qu{economical} in some sense. If one accepts the Bayesian interpretation of probability, a very powerful arsenal of inference tools become available. In a nutshell, the idea is to use the laws of probability to calculate the probabilities of different hypotheses or models, when conditioned on some known (or presumed) information.

If the collected data is denoted by $\mathbf{D}$ and the set of considered hypotheses or models is $ M_1, M_2, \ldots, M_r$, Bayes' theorem gives the plausibilities of each of the hypotheses after considering the data, the \emph{posterior probabilities},
\begin{equation}\label{eq:Bayes_model} \Pr(M_i|\mathbf{D}) = \frac{\Pr(\mathbf{D}|M_i)\Pr(M_i)}{\Pr(\mathbf{D})}.
\end{equation} 
Here, the \emph{evidence} $\Pr(\mathbf{D}|M_i)$ is the probability of the data, assuming the model $M_i$ to be true, while $\Pr(M_i)$ is the \emph{prior probability} of $M_i$, which is how plausible $M_i$ is before considering the data. $\Pr(\mathbf{D})$ is the probability of the data without assuming any particular model.\footnote{However, we note that all probabilities are always conditioned on some \qu{background information} \cite{Jaynes:book,Sivia:1996,Loredo:1990}.} One can then perform \emph{model comparison} by calculating ratios of posterior probabilities, the \emph{posterior odds}, of two models as
\begin{equation}\label{eq:post_ratio} \frac{ \Pr(M_i|\mathbf{D})}{\Pr(M_j|\mathbf{D})} =
\frac{\Pr(\mathbf{D}|M_i)}{\Pr(\mathbf{D}|M_j)} \frac{\Pr(M_i)}{\Pr(M_j)}. 
\end{equation}
In words, the posterior odds is given by the \emph{prior odds} $ \Pr(M_i)/ \Pr(M_j)$ multiplied by the \emph{Bayes factor} $B_{ij} = \Pr(\mathbf{D}|M_i)/{\Pr(\mathbf{D}|M_j)}$, which quantifies how much better $M_i$ describes that data than $M_j$.
The prior odds quantifies how much more plausible one model is than the other a priori, i.e., without considering the data. If there is no reason to favor one of the models over the other, the prior odds should be taken to equal unity (in which case the posterior odds equals the Bayes factor), but sometimes one must consider this point more carefully. 

If the model contains free parameters $\mbTh$, the evidence is given by
\bea
\mathcal{Z} =\Pr(\mathbf{D}|M) &=&  \int \Pr(\mathbf{D},\mbTh|M)\df^N\mathbf{\Theta}\notag \\ &=& \int \Pr(\mathbf{D}|\mbTh, M) \Pr(\mathbf{\Theta}|M)\df^N\mathbf{\Theta} 
\notag \\ &=&\int{\mathcal{L}(\mathbf{\Theta})\pi(\mathbf{\Theta})}\df^N\mathbf{\Theta}.
\label{eq:Z}
\eea
Here, the \emph{likelihood function} $\mathcal{L}(\mathbf{\Theta}) \equiv \Pr(\mathbf{D}|\mathbf{\Theta}, M) $ is the probability (density) of the data $\mathbf{D}$, assuming parameter values $\mbTh$ and $\pi(\mathbf{\Theta}) \equiv \Pr(\mathbf{\Theta}|M)  $ is the prior probability (density), which should reflect how plausible different values of the parameters are, assuming the model to be correct. 
It should always be normalized, i.e., integrate to unity. The assignment of priors are probably the most discussed and controversial part of Bayesian inference. This assignment is often far from trivial, 
but constitutes a very important part of any Bayesian analysis.

One observes that the evidence is the average of the likelihood over the prior, and hence this method automatically implements a form of \emph{Occam's razor}, since in general a more predictive model with a smaller parameter space will have a larger evidence than a less predictive one, unless the latter can fit the data substantially better. 

The probabilities of the different hypotheses give the complete posterior inference on the space of models, and these have a somewhat unique and meaningful interpretation on their own. However, Bayes factors, or rather posterior odds, are usually interpreted or \qu{translated} into ordinary language using the so-called \emph{Jeffreys scale}, given in \tabref{tab:Jeffreys} (\qu{$\log$} denotes the natural logarithm).

This scale has been used in applications in cosmology and astro- and particle phsycis such as Refs.~\cite{Trotta:2008qt,Hobson:2010book,Feroz:2008wr,AbdusSalam:2009tr} (and Refs.~\cite{Bergstrom:2012yi,Bergstrom:2012nx} in neutrino physics) 
although slightly more aggressive scales have been used previously \cite{Jeffreys:1961,Kass:1995}.  Note that it is often the case that the evidence is quite dependent on the prior used, 
although the Bayes factor will generally favour the correct model once \qu{enough} data have been obtained.

\begin{table}
\begin{center}
\begin{tabular}{|c|c|c|c|}
\hline
& & &\\ [-2mm]
$|\log(\text{odds})|$ & odds & $\Pr(M_1 | \mathbf{D})$ & Interpretation \\ [2mm]
\hline
& & &\\ [-2mm]
$<1.0$ & $\lesssim 3:1$ & $\lesssim 0.75$ & Inconclusive \\
$1.0$ & $\simeq 3:1$ &  $\simeq 0.75$ & Weak evidence \\
$2.5$ & $\simeq 12:1$ & $\simeq 0.92$ & Moderate evidence \\
$5.0$ & $\simeq 150:1$ & $ \simeq 0.993$ & Strong evidence \\[2mm] 
\hline
\end{tabular}
\end{center}
\caption{\it The Jeffreys scale, often used for the interpretation of Bayes factors, odds, and model probabilities. The posterior model probabilities for the preferred model are calculated by assuming only two competing hypotheses and equal prior probabilities.}
\label{tab:Jeffreys}
\end{table}

The complete inference of the parameters within a single model is given by the posterior distribution,
\begin{equation} \label{eq:bayes} 
\Pr( \mathbf{ \Theta} | \mathbf{D},M) = \frac{\Pr(\mathbf{D}
|\mathbf{\Theta},M)\Pr(\mathbf{\Theta}|M)}
{\Pr(\mathbf{D}|M)}  = \frac{\lhood(\mbTh)\pi(\mbTh)}{\ev}.
\end{equation}
Since the evidence does not depend on the values of the parameters $\mbTh$, it is usually ignored in when estimating parameters. However, often the most interesting question does not concern the parameter values within a pre-chosen model, 
but rather which are the preferred ones by the data out of a given set of models.

The main result of Bayesian parameter inference is the posterior and its marginalized versions (usually in one or two dimensions).
However, it is also common to give point estimates such as the posterior mean or median, as well as \emph{credible intervals (regions)}, which are defined as intervals (regions) containing a certain amount of posterior probability. Note that these regions are not unique without further restrictions, just as for classical confidence intervals, and that in general they do not contain all the information that the posterior contains.

Although the reasoning and techniques used when performing model selection are often different than when estimating parameters, one can equally well consider model selection as a parameter inference problem with an additional discrete parameter denoting the model index. Hence, there is no real \qu{fundamental} difference between model selection and parameter estimation. We use \MN~\cite{Feroz:2007kg,Feroz:2008xx,Feroz:2013hea} for the evaluation of all evidences and posterior distributions in this work.

%
\boldmath
\section{Analysis of specific $U(1)$ models}
\unboldmath
\label{u1mod}

In this section, we review the general strategy to build $U(1)$ models, and recall the specific models previously 
defined and discussed in Ref.~\cite{Altarelli:2012ia}. We will consider them within the context of supersymmetry, as the holomorphicity of the superpotential simplifies the construction of the Yukawa interactions. We then show the results of the Bayesian parameter estimation and model comparison. 

\boldmath
\subsection{General features}
\unboldmath

The formulation of a model based on the $U(1)$ symmetry~\cite{Froggatt:1978nt} in the supersymmetric context is simple and elegant: 
\begin{itemize}
\item[-] The flavour symmetry acts horizontally on leptons and the charges can be written as $e^c\sim(n_1^\R,n_2^\R,0)$ for the $SU(2)_\L$ lepton singlets and as $\ell\sim(n_1^\L,n_2^\L,0)$ for the $SU(2)_\L$ lepton doublets. The third lepton charges can be set to zero as only charge differences have an impact on mass hierarchies and on mixing angles. Furthermore, it is not restrictive to assume $n_1^\R>n_2^\R>0$ in order to 
guarantee the correct ordering of the charged leptons. The Higgs fields $H_{u,d}$ are not charged under $U(1)$ to prevent flavour-violating Higgs couplings.

\item[-] Once leptons have $U(1)$ charges, the Yukawa terms are no longer invariant under the action of the flavour symmetry. To formally recover the invariance, a new scalar field (or more than one in non-minimal models) can be introduced, the flavon $\theta$, that transforms non-trivially only under $U(1)$, with charge $n_\theta$. Then, the Yukawa Lagrangian can be written as 
\beq
\begin{split}
\mathcal{L}_Y =&\,(y_e)_{ij}\,\ell_i\,H_d\, e^c_j \left(\dfrac{\theta}{\Lambda}\right)^{p_e}+\\
&+(y_\nu)_{ij}\,\dfrac{\ell_i\ell_j H_u H_u}{\Lambda_\L}\left(\dfrac{\theta}{\Lambda}\right)^{p_\nu} +\text{H.c.}
\end{split}
\label{Yukawas}
\eeq
where $\Lambda$ is the cut-off of the effective flavour theory and $\Lambda_\L$ the scale of the lepton number violation, in principle distinct from $\Lambda$. $(y_e)_{ij}$ and $(y_\nu)_{ij}$ are free parameters: for naturalness, these parameters are taken to be complex and with modulus of order 1. $p_e$ and $p_\nu$ are suitable powers of the dimensionless ratio $\theta/\Lambda$ necessary to compensate the $U(1)$ charges for each Yukawa term and therefore recover the invariance under the flavour symmetry. 
Without loss of generality, we can fix $n_\theta=-1$; consequently,  $n_1,n_2>0$ to assure that the Lagrangian expansion makes sense. 
Here and in the following, neutrino masses are described by the effective Weinberg operator, while the extension to ultra-violate completions, such as See-Saw mechanisms, is straightforward.

\item[-] Once the flavon and the Higgs fields develop non-vanishing vacuum expectation values (VEVs), the flavour and electroweak symmetries are broken and mass matrices arise from the Yukawa Lagrangian. In particular, the ratio of the flavon VEV $\mean{\theta}$ and the cut-off $\Lambda$ of the effective theory defines the expanding parameter of the theory,
\beq
\ep\equiv\dfrac{\mean{\theta}}{\Lambda}<1\,.
\eeq
A useful parametrisation for the Yukawa matrices then follows as
\beq
Y_e=F_{e^c}\,y_e\,F_{\ell}\,,\qquad\qquad
Y_\nu=F_{\ell}\,y_\nu\,F_{\ell},\
\eeq
where $F_f=\diag(\ep^{n_{f1}},\ep^{n_{f2}},\ep^{n_{f3}})$. Throughout this work, and following Ref.~\cite{Altarelli:2012ia}, the charges will be taken to be integers, since non-integer charges can always be redefined to integers as long as it is accompanied by a suitable redefinition of the parameter $\epsilon$.
\end{itemize}

\boldmath
\subsection{Specific $U(1)$ models}
\label{sec:SpecificU1}
\unboldmath

The lepton charges of those models introduced in Ref.~\cite{Altarelli:2012ia} where neutrino masses are described by the Weinberg operators, are given in the upper part of Tab.~\ref{tab-models}. In the lower part, there are two new models that have been identified as \qu{good} in the analysis of \secref{sec:generalised_models}. 

\begin{table}[h!]
\begin{center}
\begin{tabular}{|c|c|c|}
\hline
& & \\ [-2mm]
{Model}& $e_\R$ & $\ell_\L$ \\ [2mm]
\hline
& & \\ 
{Anarchy ($A$)}& (3,2,0)& (0,0,0)\\ [2mm]
{$\mu\tau$-Anarchy ($A_{\mu\tau} $)}& (3,2,0) & (1,0,0)\\ [2mm]
{Hierarchy ($H$)}& (5,3,0) & (2,1,0) \\ [2mm]
\hline
& & \\ [-2mm]
{New Anarchy ($A'$)}& (3,1,0) & (0,0,0) \\ [2mm]
{New Hierarchy ($H'$)}& (8,3,0) & (2,1,0)\\ [2mm]
\hline
\end{tabular}
\end{center}
\caption{\it Upper part: the models introduced in Ref.~\cite{Altarelli:2012ia} and their flavour charges under $U(1)$. Lower part: models identified in the more general analysis of \secref{sec:generalised_models}. The flavon charge is  $-1$, while the Higgs charge is zero.}
\label{tab-models}
\end{table}

From the lepton charges in Tab.~\ref{tab-models}, the textures for the charged leptons $Y_e$ and neutrino $Y_\nu$ Yukawa matrices are as follows:
\beq
\begin{aligned}
A:\quad 
Y_e&=\left(  \begin{matrix}  \ep^3 &    \ep^2 &   1 \\  \ep^3 &    \ep^2 &   1 \\ \ep^3 &    \ep^2 &   1 \end{matrix}\right)\,,\;
Y_\nu= \left(  \begin{matrix}  1 &    1 &   1 \\  1 &   1 &  1 \\ 1 &   1 &   1 \end{matrix}\right)\,,\\
A_{\mu\tau}:\quad 
Y_e&= \left(  \begin{matrix}  \ep^4 &    \ep^3 &   \ep \\  \ep^3 &    \ep^2 &   1 \\ \ep^3 &   \ep^2 &   1 \end{matrix}    \right)\,,\;
Y_\nu= \left(  \begin{matrix}  \ep^2 &    \ep &   \ep \\  \ep &   1 &  1 \\ \ep &   1 &   1 \end{matrix} \right)\,,  \\
H:\quad 
Y_e&= \left(  \begin{matrix}  \ep^7 &    \ep^5 &   \ep^2 \\  \ep^6 &    \ep^4 &   \ep \\ \ep^5 &   \ep^3 &   1 \end{matrix} \right)\,,\;
Y_\nu= \left(  \begin{matrix}  \ep^4 &    \ep^3 &   \ep^2 \\  \ep^3 &   \ep^2 &  \ep \\ \ep^2 &   \ep &   1 \end{matrix}     \right)\,,
\end{aligned}
\label{OldModels} 
\eeq
\beq
\begin{aligned}
A':\quad 
Y_e&=\left(  \begin{matrix}  \ep^3 &    \ep &   1 \\  \ep^3 &    \ep &   1 \\ \ep^3 &    \ep &   1 \end{matrix}\right)\,,\;
Y_\nu= \left(  \begin{matrix}  1 &    1 &   1 \\  1 &   1 &  1 \\ 1 &   1 &   1 \end{matrix}\right)\,,\\
H':\quad 
Y_e&= \left(  \begin{matrix}  \ep^{10} &    \ep^6 &   \ep^2 \\  \ep^9 &    \ep^5 &   \ep \\ \ep^8 &   \ep^4 &   1 \end{matrix} \right) ,\;
Y_\nu= \left(  \begin{matrix}  \ep^4 &    \ep^3 &   \ep^2 \\  \ep^3 &   \ep^2 &  \ep \\ \ep^2 &   \ep &   1 \end{matrix}     \right)\,.
\end{aligned}
\label{NewModels} 
\eeq
In the spirit of $U(1)$ models, the coefficients in front of $\ep^n$ are expected to be complex numbers with absolute values of ${\cal O}(1)$ and arbitrary phases.
Considering that $Y_\nu$ is a symmetric matrix, the total number of parameters that should be consider in the analysis is $30$, from the Yukawa matrices, plus the unknown value of $\ep$.

In Ref.~\cite{Altarelli:2012ia}, the performances of the first three models, $A$, $A_{\mu\tau}$ and $H$, were evaluated by considering the fraction of the corresponding parameter spaces which were consistent (at a fixed confidence level) with the experimental constraints, for fixed values of $\epsilon$.  The main result of that analysis was that, using a uniform distribution for the ${\cal O}(1)$ coefficients in the interval $[0.5,2]$ or $[0.8,1.2]$ (and phases with a uniform distribution in $[0,2 \pi]$), $H$ was the best performing model for values of $\ep$ larger than about $0.3$, while for smaller values $A_{\mu\tau}$ had the best success rate. At the same time, $A$ seemed disfavoured when compared to the previous models, for almost all the range of values of $\ep$. However, it is worth to mention that i) the doublet charges of $H$ and $A_{\mu\tau}$ were chosen in order to naturally reproduce the neutrino data; ii) these models had an intrinsic advantage over $A$, as they have an extra parameter in the form of $\ep$, while $A$ is insensitive to its value; iii) although the charges and the relevant values of $\ep$ considered were chosen using the observed charged lepton masses (as well as quark masses and mixings as these models where formulated in a $SU(5)$ GUT context), they were not subsequently used in the numerical analysis.

The charges of the remaining models, $A'$ and $H'$, are chosen following the analysis in \secref{sec:generalised_models}, 
where they are treated in general as free parameters. In particular, the charges of $A'$  are identified as the best 
(i.e. having large posterior probabilities) in the case of vanishing doublet charges, whereas those 
of $H'$ are the best assuming all free charges. They are rather similar to those of $A$ and $H$, respectively.

In the remaining part of this section, we will consider the models listed in Tab.~\ref{tab-models} with the aim of:
\begin{itemize}
\item[-] Analyse and compare the models using Bayesian inference;
\item[-] Check whether the results of Ref.~\cite{Altarelli:2012ia} remain valid when performing the full Bayesian analysis (assuming that the differences in the data sets used for our analysis and that in Ref.~\cite{Altarelli:2012ia} 
are irrelevant for the model comparison); 
\item[-] Determine the importance of including the charged lepton data.
\end{itemize}

\subsection{Bayesian analysis and priors}
In order to calculate the evidence and obtain the posterior distributions, we need to specify priors on the $31$ free parameters of these models.
\begin{itemize}
\item[-] It is reasonable to take $\ep$ as a priori independent of all the $\cO(1)$ coefficients, phases, and charges, and then we use consistency of the lagrangian expansion to set an upper bound of $0.6$. We use a prior
\beq 
\pi(\epsilon) = \frac{N^{-1}(\epsilon_{0})}{1+\epsilon/\epsilon_{0}},\qquad \qquad\epsilon \in [0,0.6]\,,
\label{PriorEpsilon}
\eeq
which behaves uniformly in $\log \epsilon$ for $\epsilon \gg \epsilon_{0}$ and uniformly in $\epsilon$ for $\epsilon \ll \epsilon_{0}$. $N(\epsilon_{0})$ is the required normalisation factor. 
We take $\epsilon_{0} = 10^{-2}$ as our default choice, but we will find that our results are quite insensitive to changes in $\epsilon_{0}$, which is to be expected since it is a free parameter in all models we consider.

\item[-]  For the ${\cal O}(1)$ parameters, we also make the reasonable assumption that they are a priori independent of the charges, and so their priors should be the same for the cases of vanishing and non-zero charges.
This translates into a \emph{a priori} invariance under basis rotations and leads to a unique measure on the leptonic 
mixing matrix, the so-called Haar measure. This has been studied in some detail in 
Refs.~\cite{Hall:1999sn,Haba:2000be,deGouvea:2003xe,deGouvea:2012ac}, and 
its interpretation has often been that it describes how matrix elements or mixing angles are \qu{randomly distributed} in some sense. However, these distributions are more naturally considered in a Bayesian context, and the use of Haar measures to construct prior distributions has been extensively studied in the statistics literature (see \refcite{Kass:1996} and references therein). 

In addition to the mixing angles, there still remains an arbitrary measure over the neutrino mass eigenvalues. As pointed out in \refcite{Bai:2012zn}, under the additional assumption that the matrix elements are independent of each other, the measure on the mass matrix becomes unique (up to a scale),
\beq 
\pi(m_{ij}) \propto e^{-\Tr(m m^\dagger)/2}\,,
\eeq
and so the real and imaginary parts of each element are a priori independent with Gaussian priors. Note that, since the mass matrix is symmetric, the off-diagonal elements are on average a factor $\sqrt{2}$ smaller than the diagonal elements, but this will have a negligible impact on the results. Equivalently, the prior on the absolute value $q$ and phase $\phi$ of each (off-diagonal) element is
\beq 
\pi(q,\phi) =  \frac{q e^{-q^2/2}}{2\pi}\,, 
\label{eq:mag_prior} 
\eeq
with mode at $q=1$. 

By analogy, we take the same prior for the elements of the charged lepton mass matrix, although, since it is not symmetric, the elements have the same widths of their priors. 

\end{itemize}

\subsection{Data}
We consider the following relevant data:
\begin{itemize}
\item[-] Neutrino oscillation data constrain the parameters $ r= \Delta m_{21}^2/\Delta m_{31}^2$, $s_{12}^2$, $s_{23}^2$, and $s_{13}^2$ (using $s_{ij}$ for $\sin\theta_{ij}$). Although there are some constraints on the CP-violating phase $\delta$, we will not take this into account as induced priors on $\delta$ are rather similar and independent of $\ep$ in all models, and this will have negligible impact on the results. 
The oscillation parameters are rather well constrained and the correlations between the oscillation in the standard parameterization are 
rather small, and so we can approximate the oscillation likelihood as
\beq 
\lhood_{\rm osc}(\mbTh) \simeq \lhood^1(r)\lhood^2(s_{12}^2)\lhood^3(s_{23}^2)\lhood^4(s_{13}^2). 
\eeq
We take the individual likelihood components as Gaussian functions using the results of \refcite{GonzalezGarcia:2012sz}. Although this might not be a perfect approximation, this should not have a noticeable impact on any results.

\item[-] The ratio of the charged lepton masses, ${m_\tau}/{m_e} \simeq 3477$ and ${m_\mu}/{m_e} \simeq 207$ have been measured with very good accuracy, 
and so we can approximate the associated likelihoods with Dirac $\delta$-functions. To allow convergence of our numerical analysis we will in turn approximate these likelihoods using rather broad Gaussians. This will not change any conclusions as long as the widths of these Gaussians are taken small enough (so that all priors are effectively constant over these widths). We take our errors to be about $3-5\%$, and have checked numerically that our results are insensitive to changes in these widths.

\item[-] In what follows, only the normal mass ordering for the neutrino spectrum will be considered, 
as this is much strongly preferred than the inverted one. This is a well known result and already pointed out in Ref.~\cite{Altarelli:2002sg}: the inverted mass ordering is typically linked to a maximal value of the solar angle, in contrast with the observations.
\end{itemize}

\boldmath
\subsection{Results: constraints on $\epsilon$}\label{sec:spec_epsconstr}
\unboldmath
We first focus on the posterior distributions in all the models. However, since $\ep$ is really the only parameter of interest, and its marginal posteriors are at least not very far from being Gaussian, we only consider the posterior means and standard deviations (which closely matches the posterior medians and 68\% credible intervals).

\begin{figure}[tbh]
\includegraphics[width=0.48\textwidth]{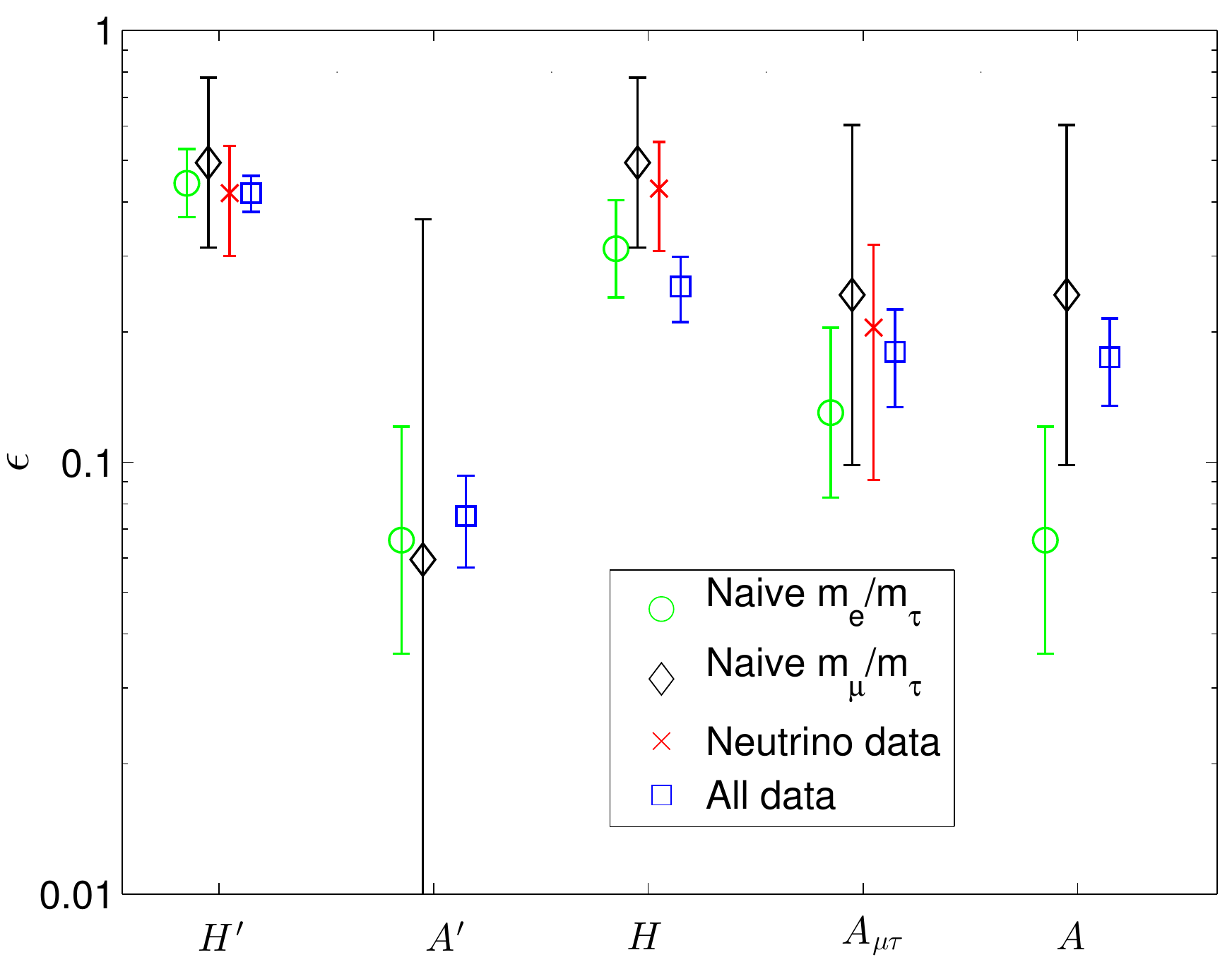}
\caption{\it Constraints on $\epsilon$ in the different models, with the error bars being twice the standard uncertainties. 
See text for the description of the naive estimates. For the numerical estimates (red and blue), the points are the posterior means and the uncertainties are the posterior standard deviations.}
\label{fig:epscons}
\end{figure}

The inclusion of the charged lepton data is particularly important since, by themselves, they can give strong constraints on $\epsilon$. 
Consequently, the same preferred values of $\epsilon$ must be used when fitting the neutrino data. In general, we expect naive estimates as
\beq \label{eq:naive}
\epsilon \simeq  \left( \frac{(y_e)_{33}\,m_e}{(y_e)_{11}\,m_\tau} \right)^{1/(n^\L_1+n^\R_1)} \simeq  \left( \frac{(y_e)_{33}\,m_\mu}{(y_e)_{22}\,m_\tau} \right)^{1/(n^\L_2+n^\R_2)},
\eeq
where $(y_e)_{ij}$ are the $\cO(1)$ factors entering the Yukawa matrix as defined in Eq.~\eqref{Yukawas}. If each of these $\cO(1)$ factors is identified with the absolute value of the matrix element in Eq.~\eqref{eq:mag_prior}, implying a prior uncertainty of $\sigma_{\log q} \simeq 0.64  $, then one can obtain naive analytical estimates of $\log \epsilon$ and its uncertainty, which can be compared with the numerical results. Note that each of the two $\cO(1)$ factors gives a contribution to the uncertainty of $\ep$. Figure \ref{fig:epscons} shows these naive estimates (in green-circles and black-diamonds, respectively) together with the numerical results when using only neutrino oscillation data (in red-crosses) and finally when also charged lepton data are considered (in blue-squares). The error bars are given by twice the naive estimates and the posterior standard deviations, respectively. In the anarchical models there are, of course, no constraints on $\epsilon$ from neutrino data.

The two naive estimates are relatively consistent with each other and with the ones using neutrino data in all the models. The largest tension can be found in the model $A$ between the electron and muon estimates, but it is still naively smaller than 3$\sigma$.\footnote{When estimating $\epsilon$ within this model, these discrepant data can still be consistently combined, but this tension will work against the model $A$ when it is compared to other models.} Furthermore, the constraints using all the data is generally consistent 
with being the naive combination of the three other sets of data (i.e. when treated as independent measurements), for both the best 
estimates and the size of the errors. For some of the models the combined uncertainty is somewhat smaller than the naive expectation, 
and for the model $H$ the posterior mean is smaller than all the three partial estimates. However, the naive combination is expected to be valid only when all the individual constraints are Gaussian and there are no common nuisance parameters. In the present case, we do see some non-Guassian features of the posteriors, and there are 30 common nuisance parameters whose correlations can invalidate the naive combination.

Finally, we note that no constraints on $\ep$ can be obtained in a standard $\chi^2$-analysis in which the $\chi^2$ is minimized (or likelihood maximized) over the remaining parameters. All the models can fit the data equally well, i.e., perfectly, for any non-zero values of $\ep$. However, depending on the value of $\ep$, the fit would require more or less fine-tuning among the $\cO(1)$ parameters. This fine-tuning is automatically considered in the Bayesian analysis and is what yields the above constraints on $\ep$.

\subsection{Results: model comparison}

All the models have a $\chi^2$-minimum of zero and therefore a $\chi^2$-analysis can never \qu{exclude} 
any of the models, and moreover two models cannot be meaningfully compared. The different models require more or 
less fine-tuning among the $\cO(1)$ parameters, and here also of $\ep$, and this aspect is automatically considered in the 
Bayesian analysis discussed here below.

We note that if $M_c$ denotes the model with fixed charges $c = (n_1^\R,n_2^\R,n_1^\L,n_2^\L)$, then
\begin{equation} \frac{ \Pr(M_{c_1}|\mathbf{D})}{\Pr(M_{c_2}|\mathbf{D})}  = \frac{\mathcal{Z}_{c_1}}{\mathcal{Z}_{c_2}} \frac{\Pr(M_{c_1})}{\Pr(M_{c_2})}.
\end{equation}
If there is no reason that a particular set of charges is a priori more plausible than any other, one sets ${\Pr(M_{c_1})}/{\Pr(M_{c_2})} = 1$ and uses the Jeffreys scale to interpret the strength of evidence. 

\begin{figure}[tbh]
\includegraphics[width=0.48\textwidth]{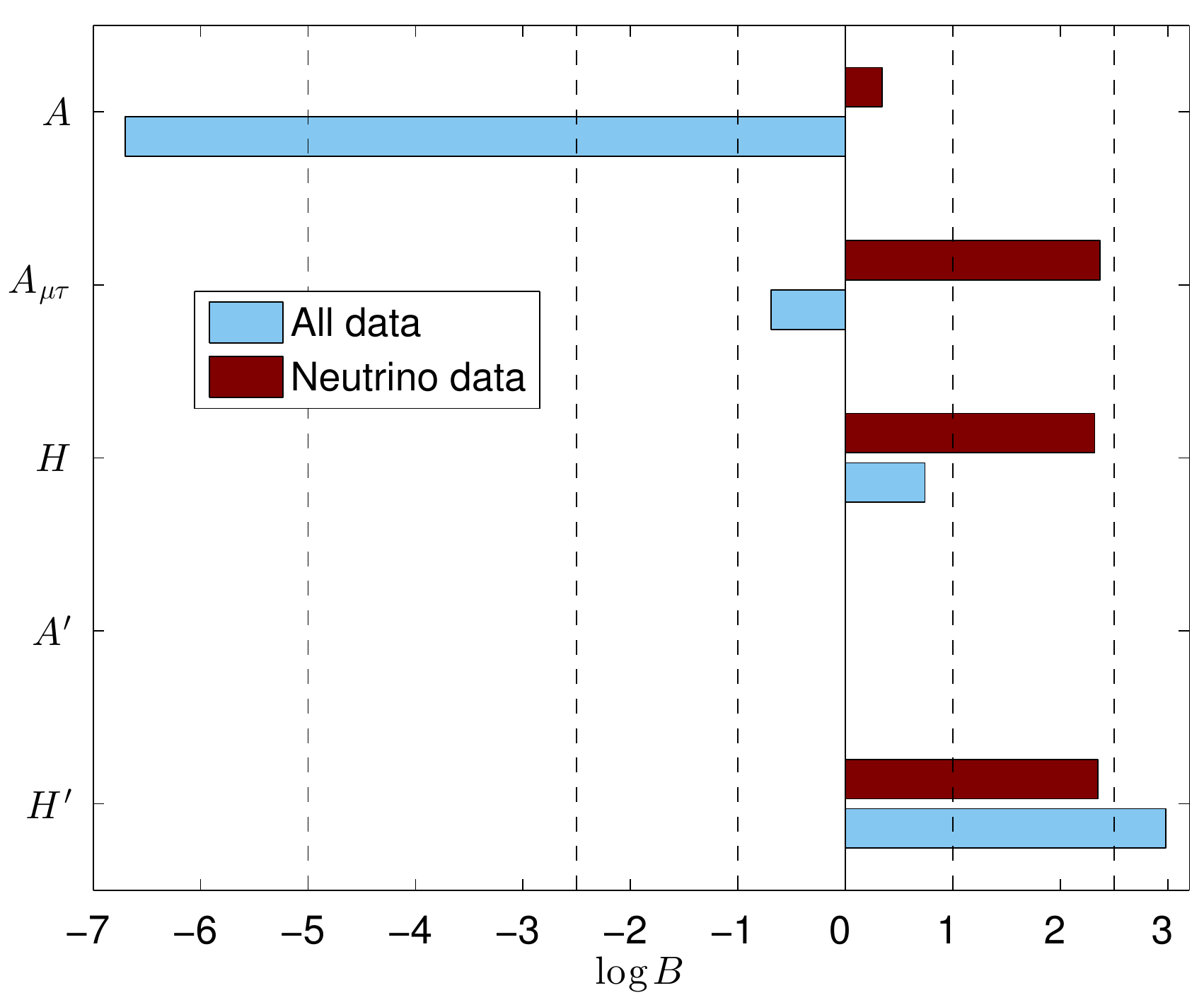}
\caption{\it Logarithms of Bayes factors with respect to the model $A'$ for the models in \tabref{tab-models} using only neutrino data (dark-red bars) and all data (light-blue bars). The dashed lines correspond to the boundaries given in \tabref{tab:Jeffreys} in the comparison with $A'$. 
}
\label{fig:Bfactors}
\end{figure}

The logarithms of the evidences of the models in \tabref{tab-models} normalised to the evidence of $A'$, i.e., the Bayes factor between all the models and $A'$, are reported in \figref{fig:Bfactors}, and this allows a quantitative comparison of all those models. The numerical (statistical) uncertainty on the individual log-evidence estimates, as reported by \MN, is about $0.15$ for all the models. Hence, the uncertainty on the logarithms of the Bayes factors are about $0.2$.  

We first focus on the models $A$, $A_{\mu\tau}$ and $H$ that have been previously discussed in Ref.~\cite{Altarelli:2012ia}. When using only the neutrino data, we see from \figref{fig:Bfactors} that the hierarchical models are all weakly preferred compared to the anarchical ones. These results are in line with the ones quoted in Ref.~\cite{Altarelli:2012ia}, in the sense that the anarchical model is less appropriate to describe the data. However, this depends somewhat on the specific choice for the prior on $\epsilon$: while the evidence of $A$ are independent on the value of $ \ep_0$, those of the hierarchical models are not. Taking, say, $\ep_0$ as small as $10^{-4}$ would reduce the $\log B$'s of $A_{\mu\tau}$ and $H$ by $0.7$ units. This can be considered the \qu{punishment} that the hierarchical models receive for their advantage of having an additional parameter, as already discussed in \secref{sec:SpecificU1}.

When also the charged lepton data is included, we notice that $A$ is strongly disfavoured compared to the other models since it predicts the charged lepton data rather badly. 

Considering in the comparison also the $H'$ and $A'$ models, $H'$ is the best model: it is moderately better 
than $A_{\mu\tau}$ and $A'$, and weakly preferred over $H$. The different evidences of $A$ and $A'$ when only neutrino data are taken into account, are due to statistical fluctuations of the evidence estimates, but the difference is consistent with being zero within the uncertainties.

The charged lepton data has again a deep impact on the analysis: once considered, $\ep$ is well-constrained in all models, which implies that there is basically no sensitivity to the volume of the prior any more, although a small dependence on the shape of the prior could still remain. 
For example, when the prior becomes effectively uniform (taking $\ep_0$ defined in Eq.~(\ref{PriorEpsilon}) very large), 
models which subsequently prefer large values of $\epsilon$ are favoured by roughly the log of the ratio of the preferred 
values of $\epsilon$'s. In this case, the evidences for $H'$ and $H$ get stronger than the values shown in \figref{fig:Bfactors}:
\beq
\begin{aligned}
H'&\rightarrow\log B \simeq 4.5\,,\\
H&\rightarrow\log B \simeq 2.5\,.
\end{aligned}
\eeq

%
%
\boldmath
\section{Generalised $U(1)$ models}
\label{sec:generalised_models}
\unboldmath

The models $A$, $A_{\mu\tau}$ and $H$ discussed in the previous section were taken as in Ref.~\cite{Altarelli:2012ia} were 
the charges have been motivated by the data. However, one could argue that in some sense the data are used 
twice - first to choose the charges and then to analyse the models. On the other hand, the charges can be thought of as just a set of 
additional unknown free parameters, which should preferably not be fixed from the beginning, but instead inferred in the Bayesian analysis. In this section, this latter strategy will be followed to study parameter constraints and make a more general comparison of anarchy vs. hierarchy.

Beside the $31$ free parameters considered when the charges are fixed, the most general model with free charges has four additional discrete parameters, which we can assign priors and subsequently calculate the posteriors of in the usual way. We consider the following generalised models:
\begin{description}
\item[${\mathcal{A}_{\text G}}$:] Generalised anarchy.  The doublets have no charges and only the $SU(2)_\L$ singlet charges are non-vanishing and are taken as free parameters.
\item[${\mathcal{H}_{\text G}}$:] Generalised hierarchy. This is the most general model where all the charges are free. Potentially, this model can give the best predictions in both the charged lepton and neutrino sectors. 
On the other hand, it has a larger parameter space and hence more regions in which it could fail to predict the experimental data. Hence, this model could as a whole have small predictive power. We include only non-vanishing doublet charges, so that ${\mathcal{A}_{\text G}}$ is not contained within setup. This will have negligible impact on our results (except of this specific region in parameter space).
\end{description}
In other words, ${\mathcal{A}_{\text G}}$ is the union of all models with anarchy in the neutrino sector, while ${\mathcal{H}_{\text G}}$ is the union of all models with hierarchy.

\subsection{Priors on the charges}
In order to perform a Bayesian analysis with free charges, one needs to assign them priors. The doublet charges should be independent on the singlet charges, but the charges of the same field must be dependent, since it holds that $n_1 \geq n_2>0$. A general prior can be written as 
\beq 
\pi(n_1,n_2) = \pi(n_1|n_2)\pi(n_2)\,.
\eeq
The charges are integers, and a naturalness criterion can be introduce such that the preference 
falls on the set with smaller charges: very large charges should not be equally plausible as small ones, but at the same time 
it is wise to not assign a prior of exactly zero to a particular charge. 
It seems reasonable to expect $\mean{n_2} = \lambda$, with $\lambda \simeq 1 - 3$.
Then the unique distribution on the non-negative reals which has maximum entropy (\qu{minimum information}, see, e.g., Refs.~\cite{Jaynes:book,Sivia:1996}) satisfying this constraint is the geometric distribution,
$ \pi(n_2) = (1-p)^{n_2}p,$ with $p=1/(\lambda+1)$. If the same condition is imposed on $n_1 - n_2$, $n_1$ is expected to be $\lambda$ larger than $n_2$ and then the prior on both charges simply becomes
\beq 
\pi(n_1,n_2) = (1-p)^{n_1}p^2\,.
\eeq
Notice that this prior only depends on the value of the largest charge and the possible dependence of the results on the value of $\lambda$ will be discussed in the next sections.

On the other side, a uniform prior $\pi(n_1,n_2)= \mathit{const.}$ for charges with $n_{\rm max}\geq n_1\geq n_2$ will also be considered.

\subsection{Results: parameter constraints}
\begin{figure}[tbh]
\includegraphics[width=0.50\textwidth,clip=true]{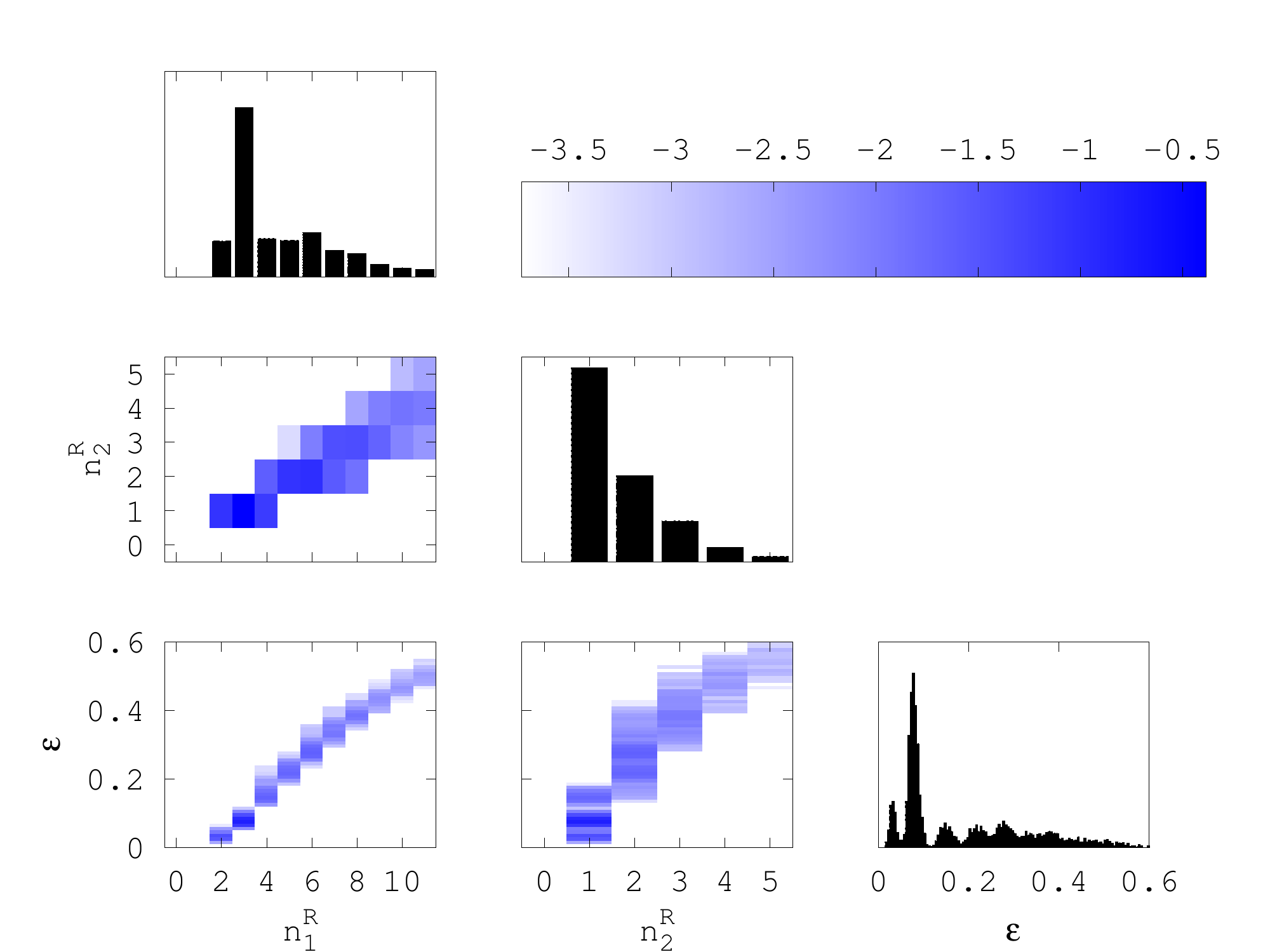}
\caption{\it Marginal posteriors of the singlet charges and $\ep$ in the model $\mathcal{A}_{\text G}$ using the full data set. In the two-dimensional plots, the color scale denotes the base-10 logarithm of the posterior probability of each region.}\label{fig:modav_an}
\end{figure}

\begin{figure*}[tbh]
\includegraphics[width=0.85\textwidth,clip=true]{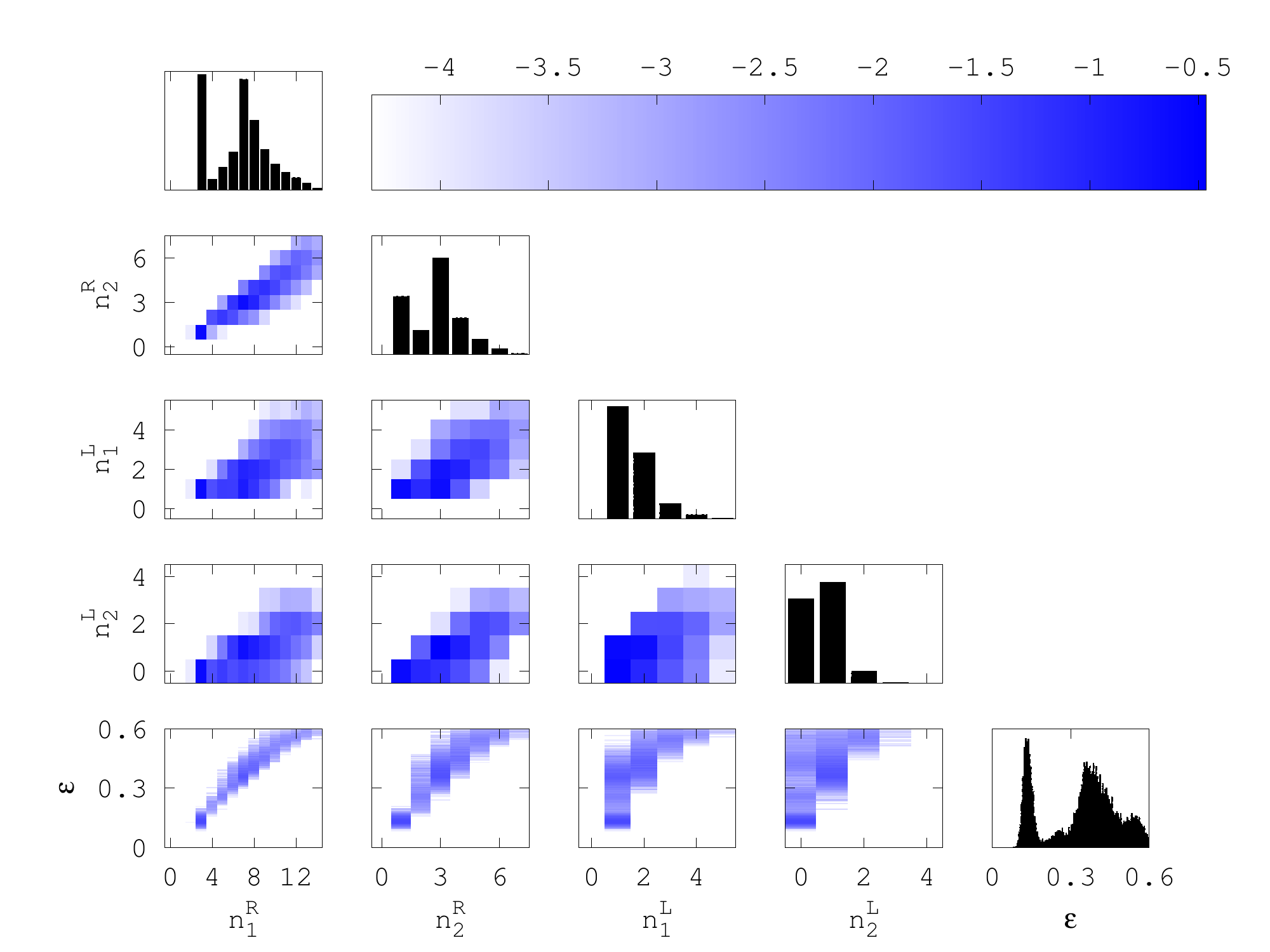}
\caption{\it Marginal posteriors of the singlet charges and $\ep$ in the model $\mathcal{M}_{\text G}$ using the full data set. In the two-dimensional plots, the color scale denotes the base-10 logarithm of the posterior probability of each region.}\label{fig:modav_hi}
\end{figure*}

It is possible now to analyse the models with free charges and compare them using the Bayesian evidence. The posteriors of the parameters of ${\mathcal{A}_{\text G}}$ and ${\mathcal{H}_{\text G}}$ models is shown in Figs.~\ref{fig:modav_an} and \ref{fig:modav_hi}, respectively.

In \figref{fig:modav_an} the posteriors of the singlet charges and $\ep$ in the generalised anarchical model is plotted using priors with $\lambda = 2$ for the mean of the charges and $\ep_0 = 10^{-2}$ for the parameter in the prior on $\ep$.\footnote{Since we find that the code becomes rather inefficient when the singlet charges are kept as free parameters when analysing this model, we evaluate the posterior by making separate runs with fixed charges and then average the results using the evidence and prior for each charge pair.} 
From the three two-dimensional posteriors of the singlet charges and $\ep$, one notes the strong degeneracy between the parameters, which follows from \eref{eq:naive} and the fact that the neutrino data is insensitive to the values of $\ep$  and the charges. Eq.~(\ref{eq:naive}) translates into $\epsilon^{n^\R_1} \simeq m_e / m_\tau , \epsilon^{n^\R_2} \simeq m_\mu / m_\tau, $ and $n^\R_1/n^\R_2 \simeq \log(m_e/m_\tau) / \log(m_\mu/m_\tau) \simeq 2.9 $. The width of these regions are determined by the widths of the $\cO(1)$ factors, and the uncertainty of $\epsilon$ is as in \secref{sec:spec_epsconstr}. 
One naively expects an exact one-dimensional degeneracy; however, this is broken by the assignment of a prior
which gives different posteriors to different points along that curve. These are the prior on the charges and on $\epsilon$, in particular its upper limit which yields upper limits also on the charges. Also the widths of the $\cO(1)$ elements affects the viability of any point along the degeneracy differently, since points which require less fine-tuning of the $\cO(1)$ elements are inherently favoured. No really strong constraints on any single parameter can be extracted though, although
with a high probability $n^\R_1  \in [2,8]$, $n^\R_2  \in [1,3]$ and $\ep<0.4$.

In the posterior of $\ep$ there are multiple peaks, each of which is generally dominated by a single pair of charges. The first three peaks correspond to charges $(2,1), (3,1)$, and  $(4,1)$, respectively, while the remaining ones which can be seen are dominated for pairs of charges with $n^\R_2 =2$.\\

In \figref{fig:modav_hi} the posteriors of all the charges and $\ep$ in the generalised hierarchical model is plotted, also using $\lambda = 2$  and $\ep_0 = 10^{-2}$. Once again, $n^\L_1 = n^\L_2 = 0$ is excluded by the definition of the model.

First, the three two-dimensional posteriors of the singlet charges and $\ep$ show that the degeneracies have changed shape and position, and gotten wider with respect to the same plots in \figref{fig:modav_an}. These are of course just what is expected from \eref{eq:naive} when the doublet charges are allowed to be non-zero. In particular, the extra suppression from the doublet charges requires $\ep$ to be larger in ${\mathcal{H}_{\text G}}$ than ${\mathcal{A}_{\text G}}$ for fixed singlet charges, while the ratio $n^\R_1/n^\R_2$ is smaller, around 2.
 
Most of the posterior probability is assigned to $(n^\L_1,n^\L_2)= (1,0),(1,1),$ and $(2,1)$.  Two of these combinations were considered in the \secref{u1mod}. We note that in general $\ep$ prefers to be larger in ${\mathcal{H}_{\text G}}$ than in ${\mathcal{A}_{\text G}}$, and also the appearance of two peaks in its posterior (on top of a large \qu{background}). From the plots of $\ep$ vs.~$n^\L_1$ and $n^\L_2$ one sees that the peak at around $\ep\approx0.15$ comes mainly from  $(1,0)$, while the peak at around $\ep\approx0.4$ is associated with $(1,1),$ and $(2,1)$. This is consistent with the fact that $(1,0)$ and $(2,1)$ were the doublet charges of the $A_{\mu \tau}$ and $H/H'$ models in \secref{sec:SpecificU1}, and that these different values of $\epsilon$ were found to be preferred in \figref{fig:epscons}.
Finally, $\ep$ has a quite strict lower bound, since for even the smallest possible choice of the doublet charges, a very small $\ep$ will give a too large hierarchy in the neutrino sector.

Furthermore, the charges for the singlets must typically be larger in $\cH_{\text G}$ than in $\cA_{\text G}$. This mainly follows from the fact that the doublet charges together with $\ep$ are essentially used to fit four observables (the three mixing angles and the ratio $r$),  while the singlet charges are used to fit only two (the mass hierarchies of the charged leptons). Then, the non-zero doublet charges require $\ep$ to not be very small so that only a mild hierarchy in the neutrino sector is achieved. Therefore, in order to fit the charged lepton data, larger singlet charges are preferred in $\cH_{\text G}$ than in $\cA_{\text G}$.

\subsection{Results: model comparison}
Similarly as in \secref{sec:SpecificU1}, no parameter constraint can be 
extracted and no model comparison can be performed from a $\chi^2$ analysis. Indeed for any given value of the charges, and indeed for any fixed tuple $(n^\R_1,n^\R_2,n^\L_1,n^\L_2,\ep)$, the data can always be fitted perfectly. 
The comparison among ${\mathcal{H}_{\text G}}$ and ${\mathcal{A}_{\text G}}$ is then performed on posteriors and evidences. 

The evidence of a model $M$ with free charges $c$ is given by
\beq 
\Pr(\mbD|M) = \Sigma_c \Pr(\mbD|c,M)\Pr(n|M) = \Sigma_c \ev_c \pi(c) 
\eeq
and so the full evidence is the average of the evidences with fixed charges over the prior on the charges.  Hence, a model with a large number of unspecified charges can have a small evidence if a 
large portion of those charges does not predict the data accurately, even if there is some specific combination of charges 
which predict the data well. 

For our default choice of priors with $\lambda = 2$ and $\ep_0 = 10^{-2}$, the comparison of ${\mathcal{H}_{\text G}}$ and ${\mathcal{A}_{\text G}}$ yields a Bayes factor
\beq 
\log \mcB = \log \left( \ev_{\mathcal{H}_{\text G}}/\ev_{\mathcal{A}_{\text G}} \right) = 1.7\,,
\eeq
which means that the hierarchical model is weakly preferred compared to the anarchical one. The uncertainty on the value reported in the previous equation, as on all the Bayes factors in this section, is about $0.2$.

Bayes factors can often depend crucially on the priors employed, in particular when the parameter in question only appears in one of the models. 
Since the doublet charges are free parameters only in $\mathcal{H}_{\text G}$, 
it is then important to check whether the results discussed above are stable under different assumptions on the priors. Usually, the prior on the additional parameters are taken as unrelated to the common parameters, while in the present case the singlet and doublet charges have the same priors, which could result in a smaller prior dependence. 

As a first possibility, the priors on the charges have been taken as the geometric distribution (as used in the default case) with the expected value $\lambda$ of the smallest charges varied between $1$ and $3$. 
A second possibility considered is of a uniform prior on the charges up to a maximum of $n_{\rm max}$ between $7$ and $15$: this implies that the marginal prior on the largest charge is proportional to its value (up to $n_{\rm max}$), and so implicitly larger charges are a priori favoured.

For the prior on $\ep$, we have varied the value of $\ep_0$, that in the previous section was fixed at $10^{-2}$. Since $\ep$ is a free parameter of all models, 
the impact on the evidence is in general expected to be small. As long as $\ep_0$ is smaller than the smallest value preferred by any model, it should not effect neither the posteriors nor the Bayes factors, and in fact all posterior inferences should have a unique limit as $\ep_0 \rightarrow 0 $ (this also applies to the specific models in \secref{sec:SpecificU1}). $\ep_0 = 10^{-2}$ is already small enough for this limit to be well approximated (we have also checked this numerically), and so we do not show any results for smaller $\ep_0$. As $\ep_0 \rightarrow \infty$, $\pi(\ep)$ becomes a uniform distribution.

We find the following.
\begin{description}
\item{\bf Geometric charge priors}\\
For small $\ep_0$ the evidence for ${\mathcal{A}_{\text G}}$ does not really depend on $\lambda$, which is expected since  in this case charges of all magnitudes are roughly equally good (along the degeneracy).
However, for $\ep_0$ with a uniform prior (i.e. large $\ep_0$), larger $\lambda$ gives larger evidence: indeed large $\epsilon$, and so larger charges, are preferred.

For ${\mathcal{H}_{\text G}}$, the effect of varying  $\lambda$ is in most cases very small, regardless of the prior on $\ep$. If only the prior on the doublet charges is made wider, the evidence decreases (since small charges predict the data better), and vice versa.
If only the prior on the singlet charges is made wider, the evidence instead increases, and vice versa. Since the priors are the same on both sets of charges, these effect partially cancel, leaving only small changes in the evidence. Since large $\ep$ is preferred in ${\mathcal{H}_{\text G}}$, the uniform prior gives a larger evidence.

When the Bayes factor is calculated, many of the changes in the evidences of the models tend to cancel, giving: 
\beq
\begin{aligned}
\log \mcB &= 1.7 \rightarrow 1.5 \quad (\lambda : 1 \rightarrow 3, ~ \ep_0 < 10^{-2} ) \\
\log \mcB &= 2.3 \rightarrow 2.0 \quad (\lambda : 1 \rightarrow 3, ~ \ep_0 = \infty  ) \,.
\end{aligned}
\eeq

Hence, in total the model comparison is very stable against changes in the priors with weak to \qu{almost moderate} preference of ${\mathcal{H}_{\text G}}$  over  ${\mathcal{A}_{\text G}}$.

\item{\bf Uniform charge priors}\\
For the uniform prior on the charges, much more prior is put on large values of the charges, especially for large $n_{\rm max}$.  Hence, one expects ${\mathcal{A}_{\text G}}$ to be relatively unaffected by this modification, while the small doublet charges required in ${\mathcal{H}_{\text G}}$ will work against it. 
Indeed, as expected, we find:
\beq
\begin{aligned}
\log \mcB &= 1.3 \rightarrow 0.8 \quad (n_{\rm max} : 7 \rightarrow 15, ~ \ep_0 < 10^{-2} ) \\
\log \mcB &= 1.8 \rightarrow 1.0 \quad (n_{\rm max} : \rightarrow 15, ~ \ep_0 = \infty  ) \,.
\end{aligned}
\eeq
\end{description}

A uniform prior in the interval $[1/3,3]$ on the  ${\cal O}(1)$ parameters has also been considered and it turns out that the difference compared to the standard (Gaussian) case is within the numerical uncertainties.

%
\section{Future prospects}
\label{Sect:Future}
In this section, we comment on the ability of future low-energy experiments to distinguish between our considered models.
If a new data set $\mbD_{\rm f}$ is added to the current one, $\mbD$, the resulting Bayes factor between two models becomes
\be \mathcal{B}_{\rm f} =  \frac{\Pr(\mbD_{\rm f},\mbD|M_1)}{\Pr(\mbD_{\rm f},\mbD|M_0)} = \frac{\Pr(\mbD_{\rm f} |\mbD, M_1)}{\Pr(\mbD_{\rm f}|\mbD, M_0)} \frac{\Pr(\mbD| M_1)}{\Pr(\mbD | M_0)}, \ee
i.e., the new Bayes factor equals the present one multiplied by $\mathcal{B}_{\rm{upd}}={\Pr(\mbD_{\rm f} |\mbD, M_1)}/{\Pr(\mbD_{\rm f}|\mbD, M_0)} $, with evidences given by
\be \Pr(\mbD_{\rm f} |\mbD, M) = \int \Pr(\mbD_{\rm f} | \mbTh, M) \Pr(\mbTh | \mbD, M) \df \mbTh .\ee
Hence, the change in the Bayes factor by the addition of a new set of data is given by calculating the average of the future likelihood over the present-day posterior. As an ideal situation, consider the case where the new data determines some combinations of parameters exactly (for example, some set of low-energy observables), so that   $\Pr(\mbD_{\rm f} | \mbTh, M) = \delta(\alpha(\mbTh)- \alpha_0)$. Then
\beq
\frac{\Pr(\mbD_{\rm f} |\mbD, M_1)}{\Pr(\mbD_{\rm f}|\mbD, M_0)} = \frac{\Pr(\alpha_0 |\mbD, M_1)}{\Pr(\alpha_0|\mbD, M_0)}, 
\eeq
and therefore if a perfect measurement of a single observable is to be able to increase the evidence of, say, 5 log units, then the ratio of the current posteriors at the true value must differ by a factor of $e^5 \simeq 150$.

Regarding the data adopted here, all the functions of parameters used to constrain the models are well-constrained, 
so that the posteriors of those parameters follow the experimental likelihoods closely in all the considered models. 
Hence, improved measurements of those parameters can not further discriminate between the models.

\begin{figure}[h!]
\includegraphics[width=0.5\textwidth]{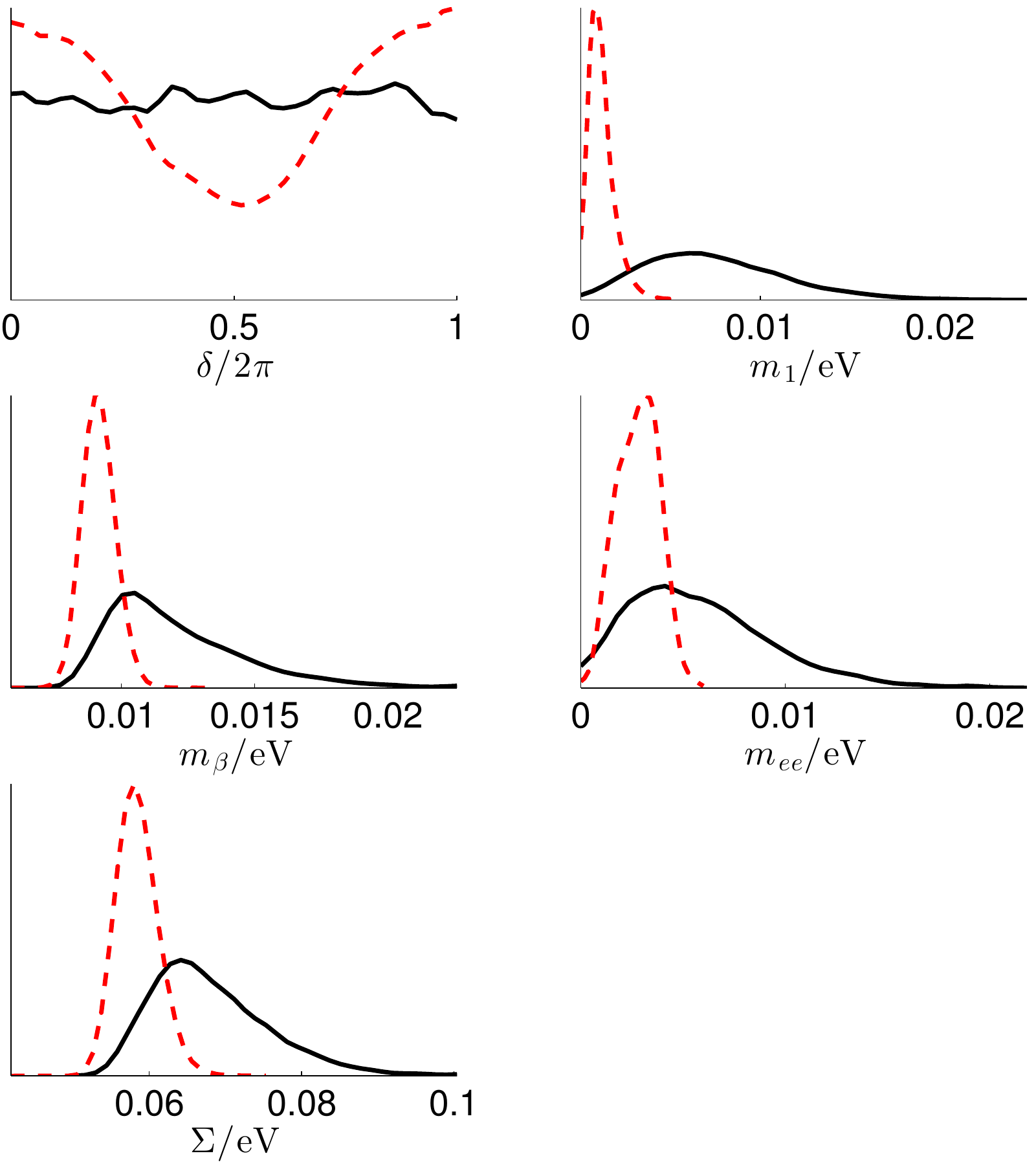}
\caption{\it Posteriors for the Dirac CP-phase $\delta$, the lightest neutrino mass $m_1$, the effective $0\nu2\beta$-decay 
effective mass $m_ {ee}$, the $\beta$-decay mass $m_\beta$, and the sum of the neutrino masses $\Sigma$, for $A'$ (black-continuous line) and $H'$ (red-dashed line).}
\label{fig:obs_1D}
\end{figure}

\begin{figure}[h!]
\includegraphics[width=0.5\textwidth]{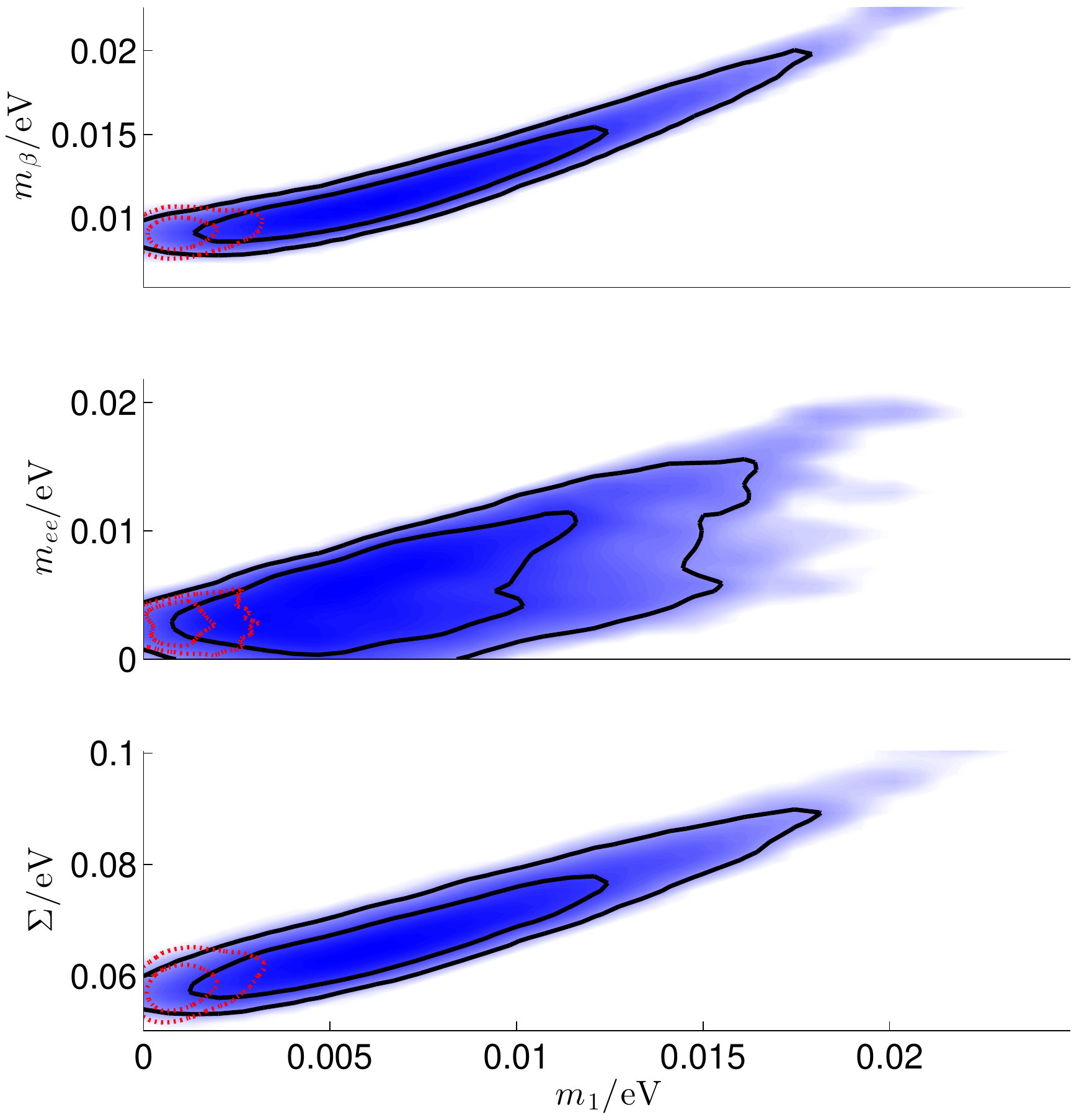}
\caption{\it The two-dimensional posterior of $m_1$ versus $m_ {ee}$, $m_\beta$, and $\Sigma$. The contours represent the $1\sigma$ and $2\sigma$ confidence level regions for $A'$ (black) and $H'$ (red).}
\label{fig:obs_2D}
\end{figure}

There are other observables which could be accurately measured in future experiments, and in principle could be used to distinguish between the models. These are primarily the CP-phase $\delta$ and observables related to the values of neutrino masses ($m_ {ee}$, $m_\beta$, $\Sigma$). We plot the posteriors of these variables as well as the lightest neutrino mass $m_1$ in \figref{fig:obs_1D} for the model $A'$ and $H'$. Similar posteriors are expected for the other hierarchical and anarchical models. Correlations among ($m_ {ee}$, $m_\beta$, $\Sigma$) and $m_1$ can be read in the two-dimensional posterior plots, displayed in \figref{fig:obs_2D}.

A precise measurement of $\delta$ can only give a very minor further discrimination between the models. 
On the other hand, a precise measurement of the sum of neutrino masses $\Sigma$ at about $0.1 \eV$ could give in principle strongly increased support for anarchical models; similar arguments hold for the other variables. 
Contrary, very stringent upper limits on the different observables could give support to the hierarchical models. 
However, the practical feasibility of these measurements are not very good in the near future. 

As discussed earlier, only the normal mass ordering for the neutrino spectrum has been considered here since it is rather strongly preferred in both the anarchical and hierarchical models. However, if definite evidence for the inverted mass ordering would emerge, all the models considered here would be disfavoured in comparison with their inverted-ordering counter parts. After that, the comparison would be between inverted hierarchical and inverted anarchical models.

%
\section{Conclusions}
\label{Conclusions}

In this letter, a conclusive comparison among anarchical and more symmetric approaches to explain the flavour puzzle in the leptonic sector is presented. The method used is based on Bayesian inference and has been applied to a series of flavour models whose effective Lagrangian shows an invariance under the Abelian group $U(1)$. Two distinct setups have been considered: i) in the first one, the $U(1)$-charges of the SM fields and of the additional scalar field responsible for the flavour symmetry breaking mechanism are fixed to specific values, determined through phenomenological considerations or by a statistical analysis; ii) in the second one, the $U(1)$-charges, or part of them, are kept as free parameters and the statistical procedure determined those ones, for which the model best reproduces the data, limiting fine-tunings. Case i) corresponds of giving to the anarchical models $A$ and $A'$
the same prior probability with respect to constructions with built-in hierarchies among the matrix elements,
$A_{\mu\tau}$, $H$ and $H'$. On the other hand, in case ii) the comparison is among the class of models embedding the anarchical ansatz $\cA_{\text G}$ and that of models based on a more symmetric principle $\cH_{\text G}$. 
In both cases, models with hierarchical matrix elements are preferred over the anarchical ones and the only difference among the two approaches is the precise values of the evidence: almost moderate in case i), and weak in case ii). This study confirms and extends the previous results in Ref.~\cite{Altarelli:2012ia}.

The stability of the results have been checked modifying the most sensible aspect of Bayesian inference, i.e. the priors on the parameters of the models. The prior of $\ep$ has been essentially taken to be either log-uniform or uniform. The priors of the $\cO(1)$ parameters entering the Yukawa matrices follow either a distribution constructed from the Haar measure or a uniform distribution in the interval $[1/3,3]$. Finally, the priors of the charges, when taken as free parameters, have chosen with either a geometric or a uniform distribution in a given support. The results show a slight dependence on the choice of priors, that however do not significantly change the conclusions.

Improvements on the precision of the used data or the addition of new data related to the neutrino mass spectrum and mixing matrix, such as $m_{ee}$, $m_\beta$, $\Sigma$ and $\delta$, will most likely a very minor effects on the results presented here. On the other hand, understanding the neutrino spectrum ordering could have an impact on the analysis: here the focus was only on the normal ordering, as the inverted one is typically linked to a maximal value of the solar angle, in contrast with the observations.

This work shows the power of the Bayesian inference in comparing models and extracting information on the model parameters. The only constructions considered here are based on the Abelian $U(1)$ flavour symmetry and with neutrino masses described by the effective Weinberg operator. Extending the analysis to the case where a See-Saw mechanism explains the lightness of the active neutrino masses is straightforward. On the other hand, to include in the comparison models based on different symmetries than $U(1)$, such us non-Abelian continuous or discrete ones, is a non trivial task and is left for future studies.

\section*{Acknowledgements}
We thank Mattias Blennow, Enrique Fernandez Martinez and Tommy Ohlsson for participating at the initial phase of this project. J.B. and L.M. acknowledge partial support by European Union FP7 ITN INVISIBLES (Marie Curie Actions, PITN-GA-2011-289442). D.M. acknowledges MIUR (Italy) for financial support under the program Futuro in Ricerca 2010 (RBFR10O36O). L.M. acknowledges partial support by the Juan de la Cierva programme (JCI-2011-09244) and by the Spanish MINECOs ``Centro de Excelencia Severo Ochoa'' Programme under grant SEV-2012-0249. L.M. thanks the G\"oran Gustafsson Foundation for financial support and KTH Royal Institute of Technology, the Dipartimento di Matematica e Fisica dell'Universit\`a Roma Tre and the theoretical group at CERN for hospitality during the development of this project.


\begin{thebibliography}{10}

\bibitem{Abe:2011sj}
{\bf T2K} Collaboration, K.~Abe {\em et.~al.},  Phys. Rev. Lett. {\bf 107}
  (2011) 041801, [\href{http://xxx.lanl.gov/abs/1106.2822}{{\tt
  arXiv:1106.2822}}].

\bibitem{Abe:2013fuq}
{\bf T2K Collaboration} Collaboration, K.~Abe {\em et.~al.},  Phys.Rev.Lett.
  {\bf 111} (2013) 211803, [\href{http://xxx.lanl.gov/abs/1308.0465}{{\tt
  arXiv:1308.0465}}].

\bibitem{Abe:2013hdq}
{\bf T2K Collaboration} Collaboration, K.~Abe {\em et.~al.},
  \href{http://xxx.lanl.gov/abs/1311.4750}{{\tt arXiv:1311.4750}}.

\bibitem{Adamson:2011qu}
{\bf MINOS} Collaboration, P.~Adamson {\em et.~al.},  Phys. Rev. Lett. {\bf
  107} (2011) 181802, [\href{http://xxx.lanl.gov/abs/1108.0015}{{\tt
  arXiv:1108.0015}}].

\bibitem{Adamson:2013ue}
{\bf MINOS Collaboration} Collaboration, P.~Adamson {\em et.~al.},
  Phys.Rev.Lett. {\bf 110} (2013), no.~17 171801,
  [\href{http://xxx.lanl.gov/abs/1301.4581}{{\tt arXiv:1301.4581}}].

\bibitem{Adamson:2013whj}
{\bf MINOS Collaboration} Collaboration, P.~Adamson {\em et.~al.},
  Phys.Rev.Lett. {\bf 110} (2013) 251801,
  [\href{http://xxx.lanl.gov/abs/1304.6335}{{\tt arXiv:1304.6335}}].

\bibitem{Abe:2011fz}
{\bf DOUBLE-CHOOZ} Collaboration, Y.~Abe {\em et.~al.},
  \href{http://xxx.lanl.gov/abs/1112.6353}{{\tt arXiv:1112.6353}}.

\bibitem{Abe:2012tg}
{\bf Double Chooz Collaboration} Collaboration, Y.~Abe {\em et.~al.},
  Phys.Rev. {\bf D86} (2012) 052008,
  [\href{http://xxx.lanl.gov/abs/1207.6632}{{\tt arXiv:1207.6632}}].

\bibitem{An:2012eh}
{\bf DAYA-BAY} Collaboration, F.~P. An {\em et.~al.},
  \href{http://xxx.lanl.gov/abs/1203.1669}{{\tt arXiv:1203.1669}}.

\bibitem{An:2013zwz}
{\bf Daya Bay Collaboration} Collaboration, F.~An {\em et.~al.},
  \href{http://xxx.lanl.gov/abs/1310.6732}{{\tt arXiv:1310.6732}}.

\bibitem{Ahn:2012nd}
{\bf RENO} Collaboration, J.~K. Ahn {\em et.~al.},
  \href{http://xxx.lanl.gov/abs/1204.0626}{{\tt arXiv:1204.0626}}.

\bibitem{Fogli:2011qn}
G.~L. Fogli, E.~Lisi, A.~Marrone, A.~Palazzo, and A.~M. Rotunno,  Phys. Rev.
  {\bf D84} (2011) 053007, [\href{http://xxx.lanl.gov/abs/1106.6028}{{\tt
  arXiv:1106.6028}}].

\bibitem{Schwetz:2011zk}
T.~Schwetz, M.~Tortola, and J.~W.~F. Valle,  New J. Phys. {\bf 13} (2011)
  109401, [\href{http://xxx.lanl.gov/abs/1108.1376}{{\tt arXiv:1108.1376}}].

\bibitem{Machado:2011ar}
P.~Machado, H.~Minakata, H.~Nunokawa, and R.~Zukanovich~Funchal,  JHEP {\bf
  1205} (2012) 023, [\href{http://xxx.lanl.gov/abs/1111.3330}{{\tt
  arXiv:1111.3330}}].

\bibitem{Tortola:2012te}
M.~Tortola, J.~Valle, and D.~Vanegas,
  \href{http://xxx.lanl.gov/abs/1205.4018}{{\tt arXiv:1205.4018}}.

\bibitem{Fogli:2012ua}
G.~Fogli, E.~Lisi, A.~Marrone, D.~Montanino, A.~Palazzo, {\em et.~al.},
  \href{http://xxx.lanl.gov/abs/1205.5254}{{\tt arXiv:1205.5254}}.

\bibitem{GonzalezGarcia:2012sz}
M.~Gonzalez-Garcia, M.~Maltoni, J.~Salvado, and T.~Schwetz,  JHEP {\bf 1212}
  (2012) 123, [\href{http://xxx.lanl.gov/abs/1209.3023}{{\tt
  arXiv:1209.3023}}]. Updates at http://www.nu-fit.org.

\bibitem{Capozzi:2013csa}
F.~Capozzi, G.~Fogli, E.~Lisi, A.~Marrone, D.~Montanino, {\em et.~al.},
  \href{http://xxx.lanl.gov/abs/1312.2878}{{\tt arXiv:1312.2878}}.

\bibitem{Altarelli:2004za}
G.~Altarelli and F.~Feruglio,  New J. Phys. {\bf 6} (2004) 106,
  [\href{http://xxx.lanl.gov/abs/hep-ph/0405048}{{\tt hep-ph/0405048}}].

\bibitem{Mohapatra:2006gs}
R.~N. Mohapatra and A.~Y. Smirnov,  Ann. Rev. Nucl. Part. Sci. {\bf 56} (2006)
  569--628, [\href{http://xxx.lanl.gov/abs/hep-ph/0603118}{{\tt
  hep-ph/0603118}}].

\bibitem{Grimus:2006nb}
W.~Grimus,  PoS {\bf P2GC} (2006) 001,
  [\href{http://xxx.lanl.gov/abs/hep-ph/0612311}{{\tt hep-ph/0612311}}].

\bibitem{GonzalezGarcia:2007ib}
M.~C. Gonzalez-Garcia and M.~Maltoni,  Phys. Rept. {\bf 460} (2008) 1--129,
  [\href{http://xxx.lanl.gov/abs/0704.1800}{{\tt arXiv:0704.1800}}].

\bibitem{Altarelli:2009wt}
G.~Altarelli,  Nuovo Cim. {\bf C32N5-6} (2009) 91--102,
  [\href{http://xxx.lanl.gov/abs/0905.3265}{{\tt arXiv:0905.3265}}].

\bibitem{Altarelli:2010gt}
G.~Altarelli and F.~Feruglio,  Rev. Mod. Phys. {\bf 82} (2010) 2701--2729,
  [\href{http://xxx.lanl.gov/abs/1002.0211}{{\tt arXiv:1002.0211}}].

\bibitem{Grimus:2011fk}
W.~Grimus and P.~O. Ludl,  \href{http://xxx.lanl.gov/abs/1110.6376}{{\tt
  arXiv:1110.6376}}.

\bibitem{Altarelli:2012ss}
G.~Altarelli, F.~Feruglio, and L.~Merlo,  Fortsch.Phys. {\bf 61} (2013)
  507--534, [\href{http://xxx.lanl.gov/abs/1205.5133}{{\tt arXiv:1205.5133}}].

\bibitem{Bazzocchi:2012st}
F.~Bazzocchi and L.~Merlo,  Fortsch.Phys. {\bf 61} (2013) 571--596,
  [\href{http://xxx.lanl.gov/abs/1205.5135}{{\tt arXiv:1205.5135}}].

\bibitem{Morisi:2012fg}
S.~Morisi and J.~Valle,  Fortsch.Phys. {\bf 61} (2013) 466--492,
  [\href{http://xxx.lanl.gov/abs/1206.6678}{{\tt arXiv:1206.6678}}].

\bibitem{King:2013eh}
S.~F. King and C.~Luhn,  Rept.Prog.Phys. {\bf 76} (2013) 056201,
  [\href{http://xxx.lanl.gov/abs/1301.1340}{{\tt arXiv:1301.1340}}].

\bibitem{Altarelli:2013eya}
G.~Altarelli,  \href{http://xxx.lanl.gov/abs/1304.5047}{{\tt arXiv:1304.5047}}.

\bibitem{King:2014nza}
S.~F. King, A.~Merle, S.~Morisi, Y.~Shimizu, and M.~Tanimoto,
  \href{http://xxx.lanl.gov/abs/1402.4271}{{\tt arXiv:1402.4271}}.

\bibitem{King:2001uz}
S.~King and G.~G. Ross,  Phys.Lett. {\bf B520} (2001) 243--253,
  [\href{http://xxx.lanl.gov/abs/hep-ph/0108112}{{\tt hep-ph/0108112}}].

\bibitem{King:2003rf}
S.~King and G.~G. Ross,  Phys.Lett. {\bf B574} (2003) 239--252,
  [\href{http://xxx.lanl.gov/abs/hep-ph/0307190}{{\tt hep-ph/0307190}}].

\bibitem{deMedeirosVarzielas:2006fc}
I.~de~Medeiros~Varzielas, S.~F. King, and G.~G. Ross,  Phys. Lett. {\bf B648}
  (2007) 201--206, [\href{http://xxx.lanl.gov/abs/hep-ph/0607045}{{\tt
  hep-ph/0607045}}].

\bibitem{Davidson:2006bd}
S.~Davidson and F.~Palorini,  Phys. Lett. {\bf B642} (2006) 72--80,
  [\href{http://xxx.lanl.gov/abs/hep-ph/0607329}{{\tt hep-ph/0607329}}].

\bibitem{Grinstein:2006cg}
B.~Grinstein, V.~Cirigliano, G.~Isidori, and M.~B. Wise,  Nucl. Phys. {\bf
  B763} (2007) 35--48, [\href{http://xxx.lanl.gov/abs/hep-ph/0608123}{{\tt
  hep-ph/0608123}}].

\bibitem{Alonso:2011jd}
R.~Alonso, G.~Isidori, L.~Merlo, L.~A. Munoz, and E.~Nardi,  JHEP {\bf 06}
  (2011) 037, [\href{http://xxx.lanl.gov/abs/1103.5461}{{\tt
  arXiv:1103.5461}}].

\bibitem{Alonso:2012fy}
R.~Alonso, M.~Gavela, D.~Hernandez, and L.~Merlo,  Phys.Lett. {\bf B715} (2012)
  194--198, [\href{http://xxx.lanl.gov/abs/1206.3167}{{\tt arXiv:1206.3167}}].

\bibitem{Alonso:2013mca}
R.~Alonso, M.~Gavela, D.~Hernández, L.~Merlo, and S.~Rigolin,  JHEP {\bf 1308}
  (2013) 069, [\href{http://xxx.lanl.gov/abs/1306.5922}{{\tt
  arXiv:1306.5922}}].

\bibitem{Alonso:2013nca}
R.~Alonso, M.~Gavela, G.~Isidori, and L.~Maiani,
  \href{http://xxx.lanl.gov/abs/1306.5927}{{\tt arXiv:1306.5927}}.

\bibitem{Altarelli:2000fu}
G.~Altarelli, F.~Feruglio, and I.~Masina,  JHEP {\bf 11} (2000) 040,
  [\href{http://xxx.lanl.gov/abs/hep-ph/0007254}{{\tt hep-ph/0007254}}].

\bibitem{Altarelli:2002sg}
G.~Altarelli, F.~Feruglio, and I.~Masina,  JHEP {\bf 01} (2003) 035,
  [\href{http://xxx.lanl.gov/abs/hep-ph/0210342}{{\tt hep-ph/0210342}}].

\bibitem{Buchmuller:2011tm}
W.~Buchmuller, V.~Domcke, and K.~Schmitz,  JHEP {\bf 03} (2012) 008,
  [\href{http://xxx.lanl.gov/abs/1111.3872}{{\tt arXiv:1111.3872}}].

\bibitem{Altarelli:2012ia}
G.~Altarelli, F.~Feruglio, I.~Masina, and L.~Merlo,  JHEP {\bf 1211} (2012)
  139, [\href{http://xxx.lanl.gov/abs/1207.0587}{{\tt arXiv:1207.0587}}].

\bibitem{Ding:2012wh}
G.-J. Ding, S.~Morisi, and J.~Valle,  Phys.Rev. {\bf D87} (2013), no.~5 053013,
  [\href{http://xxx.lanl.gov/abs/1211.6506}{{\tt arXiv:1211.6506}}].

\bibitem{Hall:1999sn}
L.~J. Hall, H.~Murayama, and N.~Weiner,  Phys. Rev. Lett. {\bf 84} (2000)
  2572--2575, [\href{http://xxx.lanl.gov/abs/hep-ph/9911341}{{\tt
  hep-ph/9911341}}].

\bibitem{Haba:2000be}
N.~Haba and H.~Murayama,  Phys. Rev. {\bf D63} (2001) 053010,
  [\href{http://xxx.lanl.gov/abs/hep-ph/0009174}{{\tt hep-ph/0009174}}].

\bibitem{deGouvea:2003xe}
A.~de~Gouvea and H.~Murayama,  Phys. Lett. {\bf B573} (2003) 94--100,
  [\href{http://xxx.lanl.gov/abs/hep-ph/0301050}{{\tt hep-ph/0301050}}].

\bibitem{deGouvea:2012ac}
A.~de~Gouvea and H.~Murayama,  \href{http://xxx.lanl.gov/abs/1204.1249}{{\tt
  arXiv:1204.1249}}.

\bibitem{Espinosa:2003qz}
J.~Espinosa,  \href{http://xxx.lanl.gov/abs/hep-ph/0306019}{{\tt
  hep-ph/0306019}}.

\bibitem{Jaynes:book}
E.~T. Jaynes, {\em {Probability theory: The logic of science }}.
\newblock Cambridge University Press, 2003.

\bibitem{Sivia:1996}
D.~S. Sivia and J.~Skilling, {\em {Data analysis: a Bayesian tutorial}}.
\newblock Oxford University Press, 2006.

\bibitem{Loredo:1990}
T.~J. Loredo,  Maximum-Entropy and Bayesian Methods (1990) 81--142.

\bibitem{Trotta:2008qt}
R.~Trotta,  Contemp.Phys. {\bf 49} (2008) 71--104,
  [\href{http://xxx.lanl.gov/abs/0803.4089}{{\tt arXiv:0803.4089}}].

\bibitem{Hobson:2010book}
M.~Hobson {\em et.~al.}, eds., {\em {Bayesian methods in cosmology}}.
\newblock Cambridge University Press, 2010.

\bibitem{Feroz:2008wr}
F.~Feroz, B.~C. Allanach, M.~Hobson, S.~S. AbdusSalam, R.~Trotta, {\em
  et.~al.},  JHEP {\bf 0810} (2008) 064,
  [\href{http://xxx.lanl.gov/abs/0807.4512}{{\tt arXiv:0807.4512}}].

\bibitem{AbdusSalam:2009tr}
S.~AbdusSalam, B.~Allanach, M.~Dolan, F.~Feroz, and M.~Hobson,  Phys. Rev. {\bf
  D80} (2009) 035017, [\href{http://xxx.lanl.gov/abs/0906.0957}{{\tt
  arXiv:0906.0957}}].

\bibitem{Bergstrom:2012yi}
J.~Bergstrom,  JHEP {\bf 1208} (2012) 163,
  [\href{http://xxx.lanl.gov/abs/1205.4404}{{\tt arXiv:1205.4404}}].

\bibitem{Bergstrom:2012nx}
J.~Bergstrom,  JHEP {\bf 1302} (2013) 093,
  [\href{http://xxx.lanl.gov/abs/1212.4484}{{\tt arXiv:1212.4484}}].

\bibitem{Jeffreys:1961}
H.~Jeffreys, {\em {Theory of probability}}.
\newblock Oxford University PRess, 1961.

\bibitem{Kass:1995}
R.~E. Kass and A.~E. Raftery,  J. Am. Stat. Ass. {\bf 90} (1995) 1773.

\bibitem{Feroz:2007kg}
F.~Feroz and M.~Hobson,  Mon. Not. Roy. Astron. Soc. {\bf 384} (2008) 449,
  [\href{http://xxx.lanl.gov/abs/0704.3704}{{\tt arXiv:0704.3704}}].

\bibitem{Feroz:2008xx}
F.~Feroz, M.~Hobson, and M.~Bridges,  Mon. Not. Roy. Astron. Soc. {\bf 398}
  (2009) 1601--1614, [\href{http://xxx.lanl.gov/abs/0809.3437}{{\tt
  arXiv:0809.3437}}].

\bibitem{Feroz:2013hea}
F.~Feroz, M.~Hobson, E.~Cameron, and A.~Pettitt,
  \href{http://xxx.lanl.gov/abs/1306.2144}{{\tt arXiv:1306.2144}}.

\bibitem{Froggatt:1978nt}
C.~D. Froggatt and H.~B. Nielsen,  Nucl. Phys. {\bf B147} (1979) 277.

\bibitem{Kass:1996}
R.~E. Kass and L.~Wasserman,  J. Am. Stat. Ass. {\bf 91} (1996) 1343.

\bibitem{Bai:2012zn}
Y.~Bai and G.~Torroba,  JHEP {\bf 1212} (2012) 026,
  [\href{http://xxx.lanl.gov/abs/1210.2394}{{\tt arXiv:1210.2394}}].

\end{thebibliography}

\providecommand{\href}[2]{#2}\begingroup\raggedright\endgroup

\end{document}